\documentclass[12pt]{article}
\usepackage{indentfirst} %% indents the first paragraph
\usepackage{amsfonts,amssymb,amsmath}
\usepackage{bm}          %% For bold math symbols
\usepackage{cite}
\usepackage{url}
\usepackage{enumitem}
\usepackage[usenames,dvipsnames]{xcolor}
\usepackage{todonotes}
\usepackage{multicol}
\usepackage{MnSymbol}
\usepackage{hyperref}
\usepackage{textcomp}
\usepackage[titletoc,title]{appendix}

\usetikzlibrary{arrows}

\AtBeginDocument{}%

\newcommand{\corurl}{blue}
\newcommand{\corcite}{ForestGreen}
\newcommand{\corlink}{blue}

\hypersetup{linktocpage,colorlinks,urlcolor=\corurl,citecolor=\corcite,linkcolor=\corlink,
pdftitle={On the distribution of the eigenvalues of the area operator in loop quantum gravity},
pdfauthor={J.F. Barbero, J. Margalef-Bentabol, E.J.S. Villaseñor}}

\numberwithin{equation}{section}  %% No need for eqnsection.sty file
\begin{document}

%\allowdisplaybreaks

\bibliographystyle{plain}

\title{On the distribution of the eigenvalues of the area operator in loop quantum gravity}

\author{
  {\small J. Fernando Barbero G.${}^{1,3}$, Juan Margalef-Bentabol${}^{1,2}$, and
          Eduardo J.S. Villase\~nor${}^{2,3}$} \\[4mm]
  {\small\it ${}^1$Instituto de Estructura de la Materia, CSIC} \\[-0.2cm]
  {\small\it Serrano 123, 28006 Madrid, Spain}         \\[1mm]
  {\small\it ${}^2$Grupo de Modelizaci\'on, Simulaci\'on Num\'erica
                   y Matem\'atica Industrial}  \\[-0.2cm]
  {\small\it Universidad Carlos III de Madrid} \\[-0.2cm]
  {\small\it Avda.\  de la Universidad 30, 28911 Legan\'es, Spain}            \\[1mm]
  {\small\it ${}^3$Grupo de Teor\'{\i}as de Campos y F\'{\i}sica
             Estad\'{\i}stica}\\[-2mm]
  {\small\it Instituto Gregorio Mill\'an, Universidad Carlos III de
             Madrid}\\[-2mm]
  {\small\it Unidad Asociada al Instituto de Estructura de la Materia, CSIC}
             \\[-2mm]
  {\small\it Madrid, Spain}           \\[-2mm]
  {\protect\makebox[5in]{\quad}}  % To force authors' names to be written
                                  %   vertically, one above another.
                                  % (\author seems to put them side-by-side
                                  %   if there is room.)
  \\
}
\date{December 19, 2017}
\maketitle
\thispagestyle{empty}   % Suppress page number on front page.

\begin{abstract}
We study the distribution of the eigenvalues of the area operator in loop quantum gravity concentrating on the part of the spectrum relevant for isolated horizons. We first show that the approximations relying on integer partitions are not sufficient to obtain the asymptotic behaviour of the eigenvalue distribution for large areas. We then develop a method, based on Laplace transforms, that provides a very accurate solution to this problem. The representation that we get is valid for any area and can be used to obtain its asymptotics in the large area limit. 
\end{abstract}


\clearpage

%%%%%%%%%%%%%%%%%%%%%%%%%%%%%%%%%%%%%%%%%%%%%%%%%%%%%%%%%%%%%%%%%%
%
% INTRODUCTION
%
\section{Introduction}{\label{sec_intro}}

The problem of understanding the properties of the spectrum of the area operator in loop quantum gravity (LQG)---in particular the distribution of its eigenvalues---has been considered by a number of authors \cite{Rovelli,Barreira,AshtLewArea,nos1}. A motivation behind some of these works was to see if the old suggestion by Bekenstein and Mukhanov regarding the quantization of black hole areas \cite{bekenstein1,Mukhanov1,BekMukh} fitted within the context of LQG, where a rigorously defined area operator, with a discrete spectrum, was constructed \cite{AshtLewArea}. Of course, it is now well known that the distribution of the eigenvalues of this operator does not conform to the simple proposal of \cite{bekenstein1,Mukhanov1,BekMukh}. In fact, there is a widespread agreement on the exponential growth of their density as a function of the area \cite{AshtLewRev,AshtLewArea}. This is supported by evidence gleaned from the observed behaviour of the lowest part of the spectrum and plausibility arguments relying on classical results about integer partitions \cite{Barreira}. However, to our knowledge, there are no definitive quantitative results on this issue. The purpose of this paper is to fill this gap and discuss in some detail the methods that, in our opinion, are best suited to address this problem.

The full spectrum of the area operator $\hat{A}_S$ associated with a surface $S$ is quite complicated (see \cite{AshtLewArea,AshtLewRev}), however, when the graphs labeling quantum states have no edges lying within $S$ and gauge invariance is enforced the eigenvalues of $\hat{A}_S$ take the following simple form
\begin{equation}\label{AreaSpectrumSimple}
   A_S=8\pi\gamma\ell_P^2\sum_{i=1}^n \sqrt{j_i(j_i+1)}\,,\quad j_i\in\tfrac{1}{2}\mathbb{N}\,,\quad  n\in\mathbb{N}\,,
\end{equation}
where $\ell_P$ denotes that Planck length and $\gamma$ is the Immirzi parameter. A context where this particular expression plays an important role is in the modelling of quantum black holes with the help of isolated horizons \cite{ABCK,ABK}.

In the following we choose units such that $4\pi\gamma\ell_P^2=1$ and introduce positive integer labels $n_i=2j_i$. By doing this the last expression becomes
\begin{equation}\label{area_spectrum}
   A_S=\sum_{i=1}^n \sqrt{n_i(n_i+2)}\,,\quad n_i\in \mathbb{N}\,,\quad  n\in\mathbb{N}\,.
\end{equation}

A natural way to study the distribution of the area eigenvalues consists in counting all the possible multisets of positive integers $n_i\in \mathbb{N}$ (maybe repeated) such that the sum in \eqref{area_spectrum} is smaller than a given value $a>0$ (for concreteness the problem is precisely spelled out in the next section).

The main difficulty to solve this problem originates in the presence of the square root. A reasonable approach to gain some preliminary understanding  on it is to consider approximations in which the square root is replaced by an integer (one expects them to be accurate if the numbers involved are large). The two most natural ones are $\sqrt{n_i(n_i+2)}\sim (n_i+1)$ and $\sqrt{n_i(n_i+2)}\sim n_i$. Approximations of this type have been the starting point of the work discussed in several papers \cite{Barreira,Sahlmann}. When these simplifications are enforced, the spectrum becomes equally spaced and essentially corresponds to the one proposed by Bekenstein and Mukhanov.

\begin{figure}
  \centering
  \includegraphics[width=\textwidth]{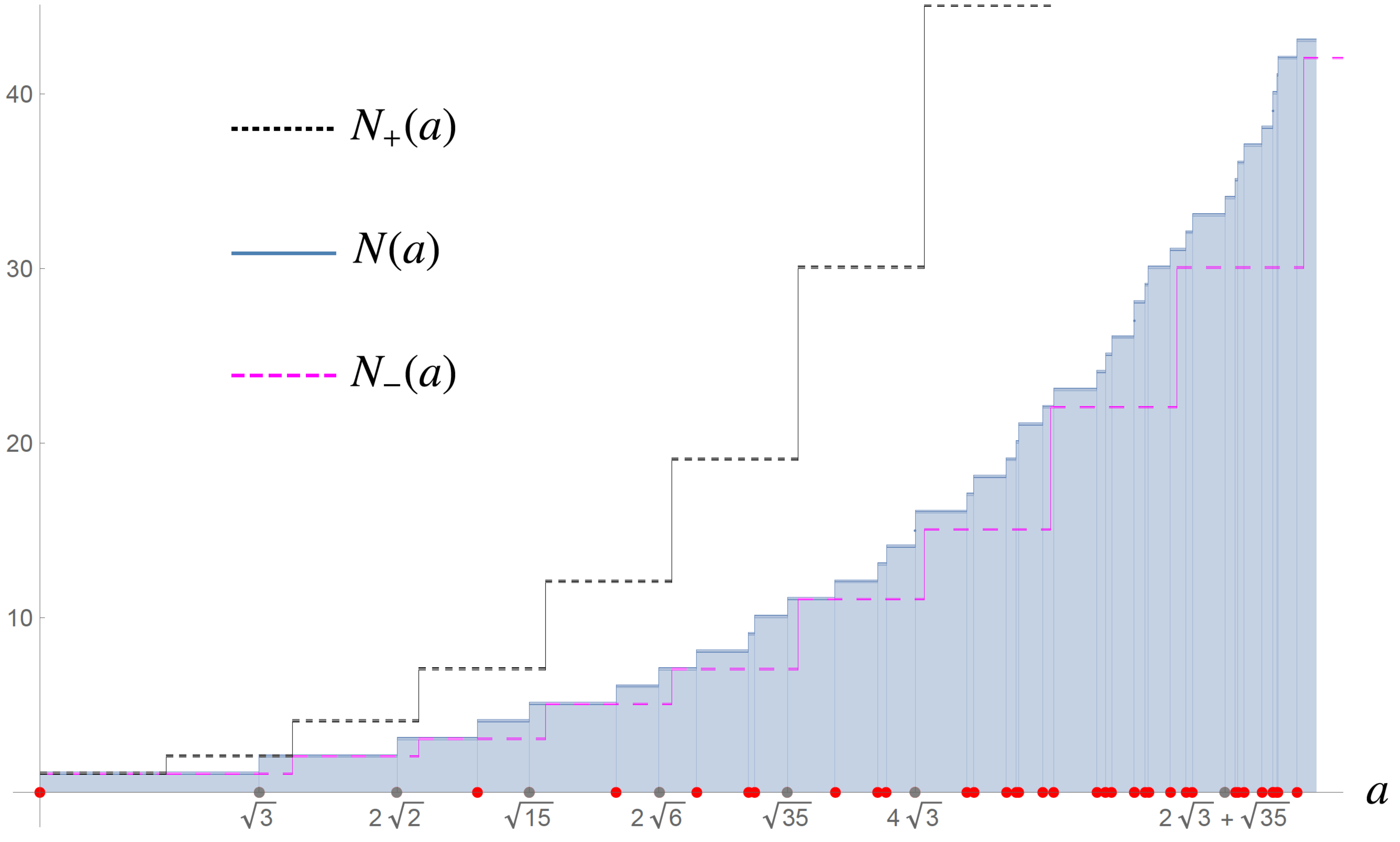}
  \caption{In this plot we compare $N(a)$, the total number of area eigenvalues smaller than an area $a$ (up to $a\sim10)$ with the two integer approximations mentioned in the text. As we can see $N_-(a)$ is not too bad for small areas but $N_+(a)$ grows too fast.}\label{fig1}
\end{figure}

For each area value $a$ let us call $N_-(a)$, $N(a)$ and $N_+(a)$ the number of different (unordered) choices of integers $n_i$ such that
\begin{equation*}
  \sum_i (n_i+1)\leq a\,,\qquad \sum_i \sqrt{n_i(n_i+2)}\leq a\,,\qquad \sum_i n_i\leq a\,,
\end{equation*}
respectively. As $n_i<\sqrt{n_i(n_i+2)}<n_i+1$ we immediately see that $N_-(a)\leq N(a)\leq N_+(a)$. A comparison between the two approximations and the exact values of $N(a)$ can be seen in figure \ref{fig1}. As it is apparent, $N_+(a)$ provides a very poor approximation for $N(a)$; it is not clear at all that the growth of $N(a)$ is captured by the one of $N_+(a)$. The behaviour of $N_-(a)$ is better but, even for the very restricted part of the spectrum shown in the figure, it is clear that $N_-(a)$ grows too slowly, raising again reasonable doubts about the possibility of capturing the behaviour of $N(a)$ with $N_-(a)$ in the large area limit.

The purpose of this paper is to understand the behaviour of $N(a)$ and assess the validity of the approximations customarily used in this setting. We will not only manage to satisfactorily do so but, along the way, we will get very accurate formulas---valid for the whole range of areas---to count the number of eigenvalues as a function of the area $a$.  The methods used here rely on well known results in analytic number theory and asymptotic analysis but introduce some novel elements. As we will justify in the paper, the standard techniques applied in the derivation of the Hardy-Ramanujan \cite{HardyRamanujan} and Rademacher \cite{Rademacher} formulas (relevant in the analysis of the approximations mentioned above) cannot be directly generalized to the counting problem that we consider here. This is so because standard generating functions cannot be used in a straightforward way.

The layout of the paper is the following. After this introduction, we devote the very short section \ref{sec_statement} to the precise statement of the counting problem that we solve. Section \ref{sec_Equally spaced} describes the integer approximations mentioned above and some generalizations of them. We will show that it is not possible to get definitive conclusions from them about the true behaviour of the area spectrum (for instance for large areas). The main body of the paper is contained in section \ref{sec_area_LQG}, where we study in detail the distribution of the eigenvalues of the area spectrum given by \eqref{area_spectrum}. We derive there a very good explicit formula that accounts for the behaviour of the \textit{whole} area spectrum. We end with our conclusions and an appendix where we show how our methods can be used to study integer partitions and obtain asymptotic expansions of the Hardy-Ramanujan and Rademacher types.

%%%%%%%%%%%%%%%%%%%%%%%%%%%%%%%%%%%%%%%%%%%%%%%%%%%%%%%%%%%%%%%%%%
%
% Statement of the problem
%
\section{Statement of the problem}{\label{sec_statement}}

For future reference and concreteness we state here the counting problem that we solve in the paper.

\bigskip

\noindent \textbf{Main Problem (MP):} \textit{For any given positive number $a>0$ compute $N(a)$ defined as one plus the number of different multisets consisting of positive integers $n_i\in \mathbb{N}$ such that
\begin{equation}\label{problem}
  \sum_{i} \sqrt{n_i(n_i+2)} \leq a\,.
\end{equation}}

\medskip

We add one for convenience (in any case, notice that zero is an eigenvalue of $\hat{A}_S$ associated with quantum geometry states labeled by graphs that do not intersect $S$). Notice that $N(a)$ is a staircase function.

An important comment is in order now: our purpose is to study the distribution of the area eigenvalues \textit{without taking into account their multiplicities} (which are relevant, on the other hand, in black hole entropy computations \cite{ABK,DomagalaLew,Meissner,nos2}). This notwithstanding, there is a ``mild'' type of multiplicity (associated with what we call $j$-degeneracy) that we will allow. It corresponds to the possibility of having distinct multisets of integers $n_i$ giving the same value for the sum $\sum_{i} \sqrt{n_i(n_i+2)}$ (for instance $\{6\}$ and $\{1,1,1,1\}$). Our choice is motivated by the fact that, on one hand, the set of degenerate eigenvalues in this sense appears to be small and, on the other, removing them complicates the computations without providing significant new insights on the problem.

%%%%%%%%%%%%%%%%%%%%%%%%%%%%%%%%%%%%%%%%%%%%%%%%%%%%%%%%%%%%%%%%%%
%
% Density of states Hardy Ramanujan
%
\section{Density of states for the integer approximations of the area spectrum}{\label{sec_Equally spaced}}

As we have mentioned in the introduction, a possible way to approach the study of the distribution of the area eigenvalues is to rely on integer approximations for the square root that appears in \eqref{area_spectrum}. We introduce now two auxiliary problems, based in approximations of this type, that we will briefly discuss in the following:

\medskip

\noindent \textbf{Auxiliary Problem 1 (AP1):} \textit{For any given positive number $a>0$ compute $N_-(a)$ defined as one plus the number of different multisets consisting of positive integers $n_i\in \mathbb{N}$ such that
\begin{equation}\label{problem1}
  \sum_{i} (n_i+1)\leq a\,.
\end{equation}}

\medskip

\noindent \textbf{Auxiliary Problem 2 (AP2):} \textit{For any given positive number $a>0$ compute $N_+(a)$ defined as one plus the number of different multisets consisting of positive integers $n_i\in \mathbb{N}$ such that
\begin{equation}\label{problem2}
  \sum_{i} n_i \leq a\,.
\end{equation}}

\medskip

Both can be conveniently rephrased in terms of integer partitions. AP1 is equivalent to computing
\[N_-(a)=1+\sum_{k\leq a}p_1(k)=1+\sum_{k=1}^{\lfloor a\rfloor}p_1(k)\,,\]
where $p_1(k)$ denotes the number of partitions of $k$ in terms of positive integers excluding 1. AP2 can be solved by computing
\[N_+(a)=1+\sum_{k\leq a}p(k)=1+\sum_{k=1}^{\lfloor a\rfloor}p(k)\,,\]
where $p(k)$ is the number of ordinary partitions of $k$ in terms of positive integers.

A neat way to encode the solutions to the two preceding counting problems is through the use of generating functions. The generating functions for $p_1(k)$ and $p(k)$ are, respectively,
\begin{equation}\label{generating_integers}
f_-(z):=\sum_{k=0}^\infty p_1(k)z^k=\prod_{m=2}^\infty \frac{1}{1-z^m}\,,\quad f_+(z):=\sum_{k=0}^\infty p(k)z^k=\prod_{m=1}^\infty \frac{1}{1-z^m}\,.
\end{equation}
Notice that, by convention, we take $p(0)=p_1(0)=1$.

For a given $n\in\mathbb{N}$ the generating functions for $N_-(n)$ and $N_+(n)$ can be obtained from \eqref{generating_integers} by simply multiplying by $1/(1-z)=1+z+z^2+z^3+\cdots$. In fact, in order to solve AP1 we need to consider the generating function
$$
\frac{f_-(z)}{1-z}\,,
$$
which is equal, actually, to the generating function for\textit{ ordinary partitions} $f_+(z)$. This means that\footnote{We denote the coefficient of the $z^n$ term in the Taylor expansion of a function $f$ around $z=0$ as $[z^n]f(z)$.}
$$
N_-(a)=[z^{\lfloor a\rfloor}]f_+(z)=p(\lfloor a\rfloor)\,,
$$
(notice that this is the approximation used in \cite{Barreira} although, in that paper, the authors use the approximation $\sqrt{n_i(n_i+2)}\sim n_i$). We have used this in figure \ref{fig1}.

An important classical result regarding the asymptotic approximation of $p(n)$ was obtained by Hardy and Ramanujan (and later perfected by Rademacher) \cite{HardyRamanujan,Rademacher} by using the so called \textit{circle method}. Their formula provides and infinite number of terms for the asymptotic expansion of these numbers and allows one, in principle, to exactly compute them. For most practical purposes it suffices to consider the first term in their expansion. It is important to notice that the alternative forms of the asymptotic approximations to $p(n)$ that appear in the literature  differ significantly with regard to their accuracy for small values of $n$ (hence, outside, the asymptotic regime). The simplest Hardy-Ramanujan formula gives
\begin{equation}\label{asymptpartitions}
N_-(a)\sim\frac{1}{4a\sqrt{3}}e^{\pi\sqrt{\frac{2a}{3}}}\,,\quad a\rightarrow+\infty\,.
\end{equation}
The asymptotic behaviour of $N_+(a)$ can also be read from the Hardy, Ramanujan and Rademacher formulas. As an example of how the methods that we use in the paper work for these problems we derive it in  appendix \ref{appendix1}. The result is
\begin{equation}\label{asymptpartitionsmas}
N_+(a)\sim\frac{1}{2\pi\sqrt{2a}}e^{\pi\sqrt{\frac{2a}{3}}}\,,\quad a\rightarrow+\infty\,.
\end{equation}
As we can see, these asymptotic behaviours are \textit{different}, although they both share the exponential factor. The growth of $N_+(a)$ is indeed faster than the one of $N_-(a)$ as the prefactors of the exponential term have different powers of $a$.

A possible improvement from the behaviour of $N_+(a)$ can be obtained by using the fact that $n<n+1/2<\sqrt{n(n+2)}$ for all $n\in\mathbb{N}$ and considering the following
\medskip

\noindent \textbf{Auxiliary Problem 3 (AP3):} \textit{For any given positive number $a>0$ compute $N_{1/2}(a)$ defined as one plus the number of different multisets consisting of positive integers $n_i\in \mathbb{N}$ such that
\begin{equation}\label{problem3}
  \sum_{i} \Big(n_i+\frac{1}{2}\Big)\leq a\,.
\end{equation}
This condition is equivalent to
\begin{equation}\label{problem33}
  \sum_{i} (2n_i+1)\leq 2a\,.
\end{equation}}

\medskip

If we define $p^{\mathrm{odd}}_1(n)$ as the number of partitions of a positive integer $n$ as sums of odd numbers excluding 1 we see that
$$
N_{1/2}(a)=\sum_{k=0}^{\lfloor2a\rfloor}p^{\mathrm{odd}}_1(k)\,.
$$
The generating function for $p^{\mathrm{odd}}_1(n)$ is
\begin{equation}\label{generating_odd}
  f_{1/2}(z):=\sum_{k=0}^\infty p^{\mathrm{odd}}_1(k)z^k=\prod_{m=1}^\infty\frac{1}{1-z^{2m+1}}\,,
\end{equation}
and, hence, the generating function for $N_{1/2}(n)$ is
$$
\prod_{m=0}^\infty\frac{1}{1-z^{2m+1}}=\prod_{m=1}^\infty (1+z^m)\,.
$$
We conclude then that $N_{1/2}(n)$ can be interpreted as the number of partitions of $n$ into \textit{distinct} summands. The asymptotic behaviour of these numbers is well known (see, for example, \cite[p.580]{Flajolet}) so we just borrow the result to get
\begin{equation}\label{asymptpartitiononehalf}
N_{1/2}(a)\sim\frac{1}{4\cdot3^{1/4}(2a)^{3/4}}e^{\pi\sqrt{\frac{2a}{3}}}\,,\quad a\rightarrow+\infty\,.
\end{equation}
This growth is slower than the one of $N_+(a)$ but faster than that of $N_-(a)$. As we can see we still get the same exponential factor as before. The interested reader can easily check that the inequality $n+r<\sqrt{n(n+2)}$ with $0<r<1$ (which is valid for large enough values of $n$) produces asymptotic approximations of the type \eqref{asymptpartitions},  \eqref{asymptpartitionsmas} and \eqref{asymptpartitiononehalf}, with different prefactors of the exponential term.

After reaching this point, it seems justified to believe that the actual asymptotic behaviour of $N(a)$ involves $\exp(\pi\sqrt{\frac{2a}{3}})$, but the other relevant factors cannot be guessed. Notice that, due precisely to the presence of the exponential term, the failure to pinpoint the exact nature of the prefactors implies that we can miss an exponentially large number of eigenvalues. In our opinion, it is not clear at all that the asymptotic behaviour of $N(a)$ can be determined by approximating $\sqrt{n(n+2)}$ in the ways explained above. A more effective approach is needed, we develop it in the next section.

%%%%%%%%%%%%%%%%%%%%%%%%%%%%%%%%%%

%%%%%%%%%%%%%%%%%%%%%%%%%%%%%%%%%%%%%%%%%%%%%%%%%%%%%%%%%%%%%%%%%%
%
% The parametrized scalar field in bounded regular spatial domains
%
\section{Distribution of the eigenvalues of the area operator}{\label{sec_area_LQG}}

A very convenient way to rephrase MP consists in determining the large $a$ behaviour of the staircase function $N(a)$ introduced in section \ref{sec_statement}. This type of function can be represented as an inverse Laplace transform (see \cite{nos3} for details). To understand why this is so, notice that the standard Heaviside step function for $a_0>0$ can be represented as
\begin{equation*}
  \theta(a-a_0)=\frac{1}{2\pi i}\int_{c-i\infty}^{c+i\infty}\frac{e^{(a-a_0)s}}{s}\mathrm{d}s\,,
\end{equation*}
with $c>0$. This means that, if we can encode the jumps $\beta_n$ located at $a=a_n$ of a certain staircase function $N(a)$ in an expansion of the type
\begin{equation*}
  \hat{f}(s)=\sum_{n=0}^\infty \beta_n e^{-a_n s}\,,
\end{equation*}
we can, at least formally, write\footnote{At the exact location of the jumps $a=a_n$ the integral gives just the arithmetic mean of the limits $\lim\limits_{a\rightarrow a_n^+} N(a)$ and $\lim\limits_{a\rightarrow a_n^-} N(a)$.}
\begin{equation*}
  N(a)=\frac{1}{2\pi i}\int_{c-i\infty}^{c+i\infty}\frac{e^{as}}{s} \hat{f}(s)\mathrm{d}s\,,
\end{equation*}
with the integration contour parallel to the imaginary axis and chosen in such a way that all the singularities of the integrand have real parts smaller than $c$.

For the AP1 of the preceding section we would have
\begin{equation*}
  \hat{f}_+(s):=f_+(e^{-s})=\prod_{n=1}^{\infty}\frac{1}{1-\exp{(-n s)}}=\exp\left(-\sum_{n=1}^\infty\log\Big(1-e^{-n s}\Big)\right)\,.
\end{equation*}
In the case of MP it is straightforward to see that
\begin{equation}\label{hatf}
  \hat{f}(s)=\prod_{n=1}^{\infty}\frac{1}{1-\exp{\big(-s\sqrt{n(n+2)}\big)}}=\exp\left(-\sum_{n=1}^\infty\log\Big(1-e^{-\sqrt{n(n+2)} s}\Big)\right)\,.
\end{equation}
If the analytic structure of $\hat{f}(s)$ was simple (for instance, meromorphic or with just a finite number of branching points allowing it to be defined on a cut plane) it would suffice to compute residues at the poles of the integrand (or wrap the integration contour around the cuts) to get an asymptotic expansion for $N(a)$. However this is not true for \eqref{hatf}, in fact, $\hat{f}(s)$ cannot be analytically extended to the region $\mathrm{Re}(s)<0$ so we need a different approach.

In the following we will use asymptotic methods relying on a saddle point approximation (in appendix \ref{appendix1} we do this for the partition problem). To this end it is necessary to get an appropriate representation for the function
\begin{equation*}
  \phi(s):=-\sum_{n=1}^\infty\log\Big(1-e^{-\sqrt{n(n+2)} s}\Big)
\end{equation*}
in the vicinity of the point $s=0$. A specially convenient one can be obtained by computing its Mellin transform and using the Mellin inversion formula to write
\begin{equation}\label{Mellin}
  \phi(s)=\frac{1}{2\pi i}\int_{\hat{c}-i\infty}^{\hat{c}+i\infty}s^{-t}\zeta(t+1)\Gamma(t)\sum_{n=1}^\infty\frac{1}{(n(n+2))^{t/2}}\mathrm{d}t\,,
\end{equation}
with $\hat{c}>1$. Although, in principle, the representation provided by \eqref{Mellin} is only valid for $s\in\mathbb{R}$ it can be extended for complex values of $s$ by relying on the argument given in \cite[p.576]{Flajolet}.

As we can see, the integrand in \eqref{Mellin} consists of a number of pieces: the usual $s^{-t}$, two ``universal'' factors $\Gamma(t)$ and $\zeta(t+1)$ (compare with \eqref{inverseMellinfitilde} in appendix \ref{appendix1}), and a Dirichlet series (a generalized zeta function)
\begin{equation*}
  \sum_{n=1}^\infty\frac{1}{(n(n+2))^{t/2}}\,.
\end{equation*}
This last object can be extended to the complex plane as a meromorphic function $\zeta_A(t)$ with an infinite number of isolated simple poles located at $t=1\,,-1\,,-3\,,-5\,,\ldots$ Indeed, in terms of the Hurwitz zeta function\footnote{Remember that for $z$ such that $\mathrm{Re}(z)>0$ and $\mathrm{Re}(a)>0$ the Hurwitz zeta function is defined by the series $\zeta(z,a)=\sum\limits_{n=0}^\infty(n+a)^{-z}$.} we can write
\begin{equation*}
  \zeta_A(t)=\sum_{\ell=0}^\infty 2^\ell\frac{\Gamma(\ell+t/2)}{\ell!\,\Gamma(t/2)} \zeta(t+\ell,3)\,.
\end{equation*}
With the help of this extension we can consider displacing the integration contour of \eqref{Mellin} to the left and get the sought for representation of $\phi(s)$ by computing the residues of
\begin{equation*}
  F(s,t):=s^{-t}\zeta(t+1)\Gamma(t)\zeta_A(t)
\end{equation*}
at its poles $t=+1\,,0\,,-1\,,-2\,,-3\,,\ldots$ The representation that we are about to get is valid close to $s=0$ which is \textit{precisely} what we need to obtain the asymptotic behaviour of $N(a)$ in the limit $a\rightarrow+\infty$. The residues at these poles are
\begin{align*}
    &\mathrm{Res}[F(s,t);t=+1]=\frac{\pi^2}{6s}\,, \\
    &\mathrm{Res}[F(s,t);t=0]=\frac{3}{2}\log s-\log\sqrt{\pi}\,,   \\
    &\mathrm{Res}[F(s,t);t=-1]=\frac{1}{4}s\log s-a_* s\,,  \\
    &\mathrm{Res}[F(s,t);t=-2m]=-\frac{\zeta(2m)}{2m}\left(\frac{s}{2\pi}\right)^{2m}\,,\hspace*{4.35cm} m\in\mathbb{N}\,,  \\
    &\mathrm{Res}[F(s,t);t=-2m-1]=\frac{\sqrt{\pi}}{2}\frac{\zeta(2m+1)\Gamma(m+1/2)}{(m+1)!}\left(\frac{s}{2\pi}\right)^{2m+1}\,,\quad m\in\mathbb{N} \,,
\end{align*}
where the constant $a_*$ is given by the series
\begin{align*}
  a_*:=&\frac{1}{6}+\frac{1}{4}\log(2\pi)+\frac{1}{4\sqrt{\pi}}\sum_{n=3}^\infty\frac{2^n}{n!}\Gamma\Big(n-\frac{1}{2}\Big)\zeta(n-1,3)\\
    =&\frac{1}{6}+\frac{1}{4}\log(2\pi)+\frac{1}{2}\sum_{n=3}^{\infty}n\left(1-\frac{1}{n}-\frac{1}{2n^2}-\sqrt{1-\frac{2}{n}}\right)\sim0.765115321592\cdots
\end{align*}
By adding the contribution of these residues we obtain
\begin{align}\label{expphi}
 \hat{f}(s)&=\frac{s^{3/2}}{\sqrt{\pi}}\exp\Big(\frac{\pi^2}{6s}-a_*s+\frac{1}{4}s\log s\Big)\nonumber\\
  &\cdot\exp\left[\sum_{m=1}^N\left(-\frac{\zeta(2m)}{2m}\left(\frac{s}{2\pi}\right)^{2m}\!\!\!\!+
  \frac{\sqrt{\pi}}{2}\frac{\zeta(2m+1)\Gamma(m+1/2)}{(m+1)!}\left(\frac{s}{2\pi}\right)^{2m+1}\right)\right]\\
  &\cdot\exp\big(O(s^N)\big)\,,\quad\forall N\in\mathbb{N}\,,\quad s\rightarrow0\,.\nonumber
  \end{align}
Several comments are in order now. First of all, the series appearing in the preceding expression converges whenever $|s|<2\pi$. This is enough for our purposes, however it is interesting to point out that a nice analytic extension of it can be obtained by taking into account that
\begin{align*}
  &-\sum_{m=1}^\infty\frac{\zeta(2m)}{2m}\left(\frac{s}{2\pi}\right)^{2m}=\frac{1}{2}\log\left(\frac{2}{s}\sin\frac{s}{2}\right)\,, \\
  &\exp\left[\!\frac{\sqrt{\pi}}{2}\!\sum_{m=1}^{\infty}\!\frac{\zeta(2m+1)\Gamma(m+1/2)}{(m+1)!}\left(\frac{s}{2\pi}\right)^{2m+1}\!\right]\!=\!\prod_{k=1}^{\infty}\!\exp\left[\!-\frac{\pi^2 k}{s}\left(\sqrt{\!4\!-\!\frac{s^2}{\pi^2k^2}}\!-\!2\!+\!\frac{s^2}{4\pi^2 k^2}\right)\right]\,.
\end{align*}
Another important thing to notice is the fact that $\hat{f}(s)$ is not exactly given by the terms appearing in the first two lines of \eqref{expphi} because of an extra contribution $G(s):=\exp\big(O(s^N)\big)\,,\forall N\in\mathbb{N}\,,s\rightarrow0$. This function plays no role in the final asymptotic formulas that we will get in the paper because it approaches 1 very fast as $s\rightarrow0^+$ but it would be important in order to recover the full analytic structure of $\hat{f}(s)$.

In the following we will write
\begin{align}
  F(s):&=\exp\left[\sum_{m=1}^{\infty}\left(-\frac{\zeta(2m)}{2m}\left(\frac{s}{2\pi}\right)^{2m}+\frac{\sqrt{\pi}}{2}\frac{\zeta(2m+1)\Gamma(m+1/2)}{(m+1)!}\left(\frac{s}{2\pi}\right)^{2m+1}\right)\right]\label{funcionF}\\
  &=\sqrt{\frac{2}{s}\sin\frac{s}{2}}\exp\left(\frac{\sqrt{\pi}}{2}\sum_{m=1}^\infty\frac{\zeta(2m+1)\Gamma(m+1/2)}{(m+1)!}\left(\frac{s}{2\pi}\right)^{2m+1} \right)\nonumber\,.
\end{align}

\subsection{Computation of \texorpdfstring{$N(a)$}{N(a)}: approximations and asymptotic analysis}\label{subsec_approx}

Here we will concentrate on the computation of
\begin{equation}\label{Na}
  N(a)=\frac{1}{2\pi^{3/2}i}\int_{c-i\infty}^{c+i\infty}s^{1/2}\exp\Big((a-a_*)s+\frac{\pi^2}{6s}+\frac{1}{4}s\log s\Big)F(s)G(s)\mathrm{d}s\,,
\end{equation}
where $a_*$, $F(s)$ and $G(s)$ have been defined in the previous section. We will not content ourselves with an asymptotic approximation but will try to get a representation for $N(a)$ valid for \textit{any value} of the area $a$. As we will see, the particular features of \eqref{Na} will allow us to obtain a rather precise functional representation for it, in addition to a simple asymptotic expansion in the limit $a\rightarrow+\infty.$

The first step in the determination of $N(a)$ consists in performing the change of variable $s=\alpha t$ leading to
\begin{equation}\label{Na1}
\hspace*{-2mm}N(a)=\!\!-\frac{i}{2}\left(\frac{\alpha}{\pi}\right)^{3/2}\!\!\!\!\int_{\hat{c}-i\infty}^{\hat{c}+i\infty}\!\!t^{1/2}\exp\left(\alpha\Big(a\!-\!a_*\!+\!\frac{1}{4}\log\alpha\Big)t\!+\!\frac{\pi^2}{6\alpha t}\!+\!\frac{1}{4}\alpha t\log t\right)F(\alpha t)G(\alpha t)\mathrm{d}t\,.
\end{equation}
By judiciously choosing $\alpha$ we can write the exponential term as
\begin{equation}\label{exp}
  \exp\left(\frac{\pi^2}{6\alpha}\left(e^2t+\frac{1}{t}\right)+\frac{1}{4}\alpha t\log t\right)\,.
\end{equation}
This expression is useful because the terms $e^2t+1/t$ and $t\log t$ \textit{both} have a real stationary point at $t=1/e$. As we will see, this will help us find a very good approximation for $N(a)$. Another reason why \eqref{exp} is specially useful is the way the $a$-dependent parameter $\alpha$ appears: as $1/\alpha$ multipliying the first term and as $\alpha$ multiplying the second.

The value of $\alpha$ leading to \eqref{exp} is
\begin{equation}\label{alpha}
  \alpha=\exp\left(\frac{1}{2}W\left(\frac{4\pi^2e^2}{3}e^{8(a-a_*)}\right)-4(a-a_*)\right)\,,
\end{equation}
where $W$ is the Lambert function\footnote{The Lambert function is defined by the implicit equation $W(z)\exp W(z)=z$.}. Remembering that $W(x)\sim\log x-\log(\log x)$ as $x\rightarrow+\infty$, it is straightforward to see that
\begin{equation*}
  \alpha\sim\frac{\pi e}{\sqrt{6a}}\,, \quad a\rightarrow+\infty\,.
\end{equation*}
This fact shows an additional advantage of the representation provided by \eqref{Na1}: we can use $1/\alpha$ as our asymptotic parameter instead of $a$.

Let us look again at the exponential term \eqref{exp}. Owing to the fact mentioned above regarding the stationary points of $e^2t+1/t$ and $t\log t$, we can actually approximate $t\log t$ in a neighborhood of $t=1/e$ as
\begin{equation*}
  t\log t=-\frac{2}{e}+\frac{1}{2e^2}\left(e^2t+\frac{1}{t}\right)+O\left(t-\frac{1}{e}\right)^3
\end{equation*}
so that
\begin{equation}\label{exp2}
  \exp\left(\frac{1}{4}\alpha t\log t\right)=\exp\left(-\frac{\alpha}{2e}\right)\cdot\exp\left(\frac{\alpha}{8e^2}\left(e^2t+\frac{1}{t}\right)\right)\cdot\exp\left(O\left(t-\frac{1}{e}\right)^3\right)\,,
\end{equation}
and, hence, we can get a very good--but simple--approximation for \eqref{exp}:
\begin{equation}\label{exp3}
   \exp\left(\frac{\pi^2}{6\alpha}\left(e^2t+\frac{1}{t}\right)+\frac{1}{4}\alpha t\log t\right)\sim\exp\left(-\frac{\alpha}{2e}\right)\cdot\exp\left(\left(\frac{\alpha}{8e^2}
   +\frac{\pi^2}{6\alpha}\right)\left(e^2t+\frac{1}{t}\right)\right)\,.
\end{equation}
It is the particular form in which the integration variable $t$ appears in the previous expression that will allow us to get a very accurate representation for $N(a)$ in terms of modified spherical Bessel functions.

Finally, in a neighborhood of $t=1/e$, we write $G(\alpha t)$ as $1$ and $F(\alpha t)$ as
\begin{equation}\label{Fapprox}
  F(\alpha t)=F\left(\frac{\alpha}{e}\right)\left(1-eH(\alpha)\left(t-\frac{1}{e}\right)\right)+O\left(t-\frac{1}{e}\right)^2\,,
\end{equation}
where
\begin{equation}\label{H}
  H(\alpha)=\sum_{m=1}^\infty
\left(\zeta(2m)\left(\frac{\alpha}{2\pi e}\right)^{2m}\!\!-\sqrt{\pi}\frac{\zeta(2m+1)\Gamma(m+3/2)}{(m+1)!}\left(\frac{\alpha}{2\pi e}\right)^{2m+1}\right)\,.
\end{equation}

Combining all the previous elements we can now easily write successive approximations for $N(a)$. Here we will just consider the first two ones:
\begin{align}\label{approxN}
  N_0(a) & =-\frac{i}{2}\left(\frac{\alpha}{\pi}\right)^{3/2}e^{-\frac{\alpha}{2e}}F\left(\frac{\alpha}{e}\right)\int_{\hat{c}-i\infty}^{\hat{c}+i\infty}t^{1/2}
            \exp\left(\left(\frac{\alpha}{8e^2}+\frac{\pi^2}{6\alpha}\right)\left(e^2t+\frac{1}{t}\right)\right)\mathrm{d}t\,,\\
  N_1(a) & =-\frac{i}{2}\left(\frac{\alpha}{\pi}\right)^{3/2}e^{-\frac{\alpha}{2e}}F\left(\frac{\alpha}{e}\right)(1+H(\alpha))\int_{\hat{c}-i\infty}^{\hat{c}+i\infty}t^{1/2}
            \exp\left(\left(\frac{\alpha}{8e^2}+\frac{\pi^2}{6\alpha}\right)\left(e^2t+\frac{1}{t}\right)\right)\mathrm{d}t\nonumber\\
         +&\frac{ie}{2}\left(\frac{\alpha}{\pi}\right)^{3/2}e^{-\frac{\alpha}{2e}}F\left(\frac{\alpha}{e}\right)H(\alpha)\int_{\hat{c}-i\infty}^{\hat{c}+i\infty}t^{3/2}
            \exp\left(\left(\frac{\alpha}{8e^2}+\frac{\pi^2}{6\alpha}\right)\left(e^2t+\frac{1}{t}\right)\right)\mathrm{d}t\,.
\end{align}

A very nice, almost exact formula, for these expressions---our final trick---can be obtained by realizing that the exponent in the integrand is a characteristic feature of the contour integral representation of Bessel and modified Bessel functions. Indeed, by performing the change of variable\footnote{Of course, it is possible to perform the two changes of variables that we use in this section in a single step by requiring the minimum of the exponent in \eqref{Na1} to be located at $\tau=1$. We have chosen a slighter longer presentation because it highlights the main ideas leading to the main result of the paper.} $\tau=et$ and changing the contour to a counterclockwise oriented circle $C$ centered in the origin (which introduces only exponentially suppressed subdominant terms) we get the main result of our paper
\begin{align}\label{N0}
  N_0(a)&\sim -\frac{i}{2}\left(\frac{\alpha}{\pi e}\right)^{3/2}e^{-\frac{\alpha}{2e}}F\left(\frac{\alpha}{e}\right)\int_C \tau^{1/2}
            \exp\left(\left(\frac{\alpha}{8e}+\frac{\pi^2e}{6\alpha}\right)\left(\tau+\frac{1}{\tau}\right)\right)\mathrm{d}\tau\,\nonumber\\
        &=\frac{1}{\sqrt{\pi}}\left(\frac{\alpha}{e}\right)^{3/2}e^{-\frac{\alpha}{2e}}F\left(\frac{\alpha}{e}\right)I_{-3/2}\left(\frac{\pi^2 e}{3\alpha}+\frac{\alpha}{4e}\right)\,,
\end{align}
where $I_{-3/2}$ is a modified Bessel function of the first kind
\begin{equation*}
  I_{-3/2}(z)=\sqrt{\frac{2}{\pi}}\frac{1}{z^{3/2}}\left(z\sinh z-\cosh z\right)\,.
\end{equation*}
In a similar fashion one finds
\begin{align}\label{N1}
  N_1(a)&\sim\\
  &\frac{1}{\sqrt{\pi}}\left(\frac{\alpha}{e}\right)^{3/2}e^{-\frac{\alpha}{2e}}F\left(\frac{\alpha}{e}\right)\left((1+H(\alpha))I_{-3/2}\left(\frac{\pi^2 e}{3\alpha}+\frac{\alpha}{4e}\right)-H(\alpha)I_{-5/2}\left(\frac{\pi^2 e}{3\alpha}+\frac{\alpha}{4e}\right)\right)\,,\nonumber
\end{align}
where
\begin{equation*}
  I_{-5/2}(z)=\sqrt{\frac{2}{\pi}}\frac{1}{z^{5/2}}\left((3+z^2)\cosh z-3z\sinh z\right)\,.
\end{equation*}
It is now straightforward to get from \eqref{N0} an asymptotic expansion for $N(a)$ in the limit $a\rightarrow+\infty$
\begin{equation}\label{asimpt1}
  N(a)\sim\frac{1}{2\sqrt{6}a}e^{\pi\sqrt{\frac{2a}{3}}}=:N_{\mathrm{as}}(a)\,.
\end{equation}
Although this expansion can obviously be obtained in a more direct way from \eqref{Na} by using a saddle point approximation, we think that the possibility of finding a representation such as \eqref{N0} has its merit, because it provides a very good approximation for low values of the area. We discuss this issue in the next section

%%%%%%%%%%%%%%%%%%%%%%%%%%%%%%%%%%%%%%%%%%%%%%%%%%%%%%%%%%%%%%%%%%
%
% Checking the results: conclusions and comments
%
\section{Checking the results: conclusions and comments}{\label{sec_check}}

The purpose of this section is to assess the quality of the approximations furnished by (\ref{N0}) and (\ref{N1}) both at the asymptotic regime $a\rightarrow+\infty$ and for small values of $a$. We also want to check if the behaviour predicted by \eqref{asimpt1} is compatible with the one obtained from the approximations based on the partition problem, in particular $N_-(a)$. This last question can be immediately answered; by comparing \eqref{asimpt1} and \eqref{asymptpartitions} we see that both expressions are proportional but not equal. The asymptotic approximation obtained from the partition problem and the Hardy-Ramanujan formula is smaller than  \eqref{asimpt1} by a factor of $\sqrt{2}$. This immediately explains the behaviour shown in figure \ref{fig1}. Although one can argue that both asymptotic behaviours are not dramatically different (their functional forms are essentially identical) it is important to realize that the difference between them grows exponentially.

\begin{figure}[!ht]
	\centering
	\includegraphics[width=14.9cm]{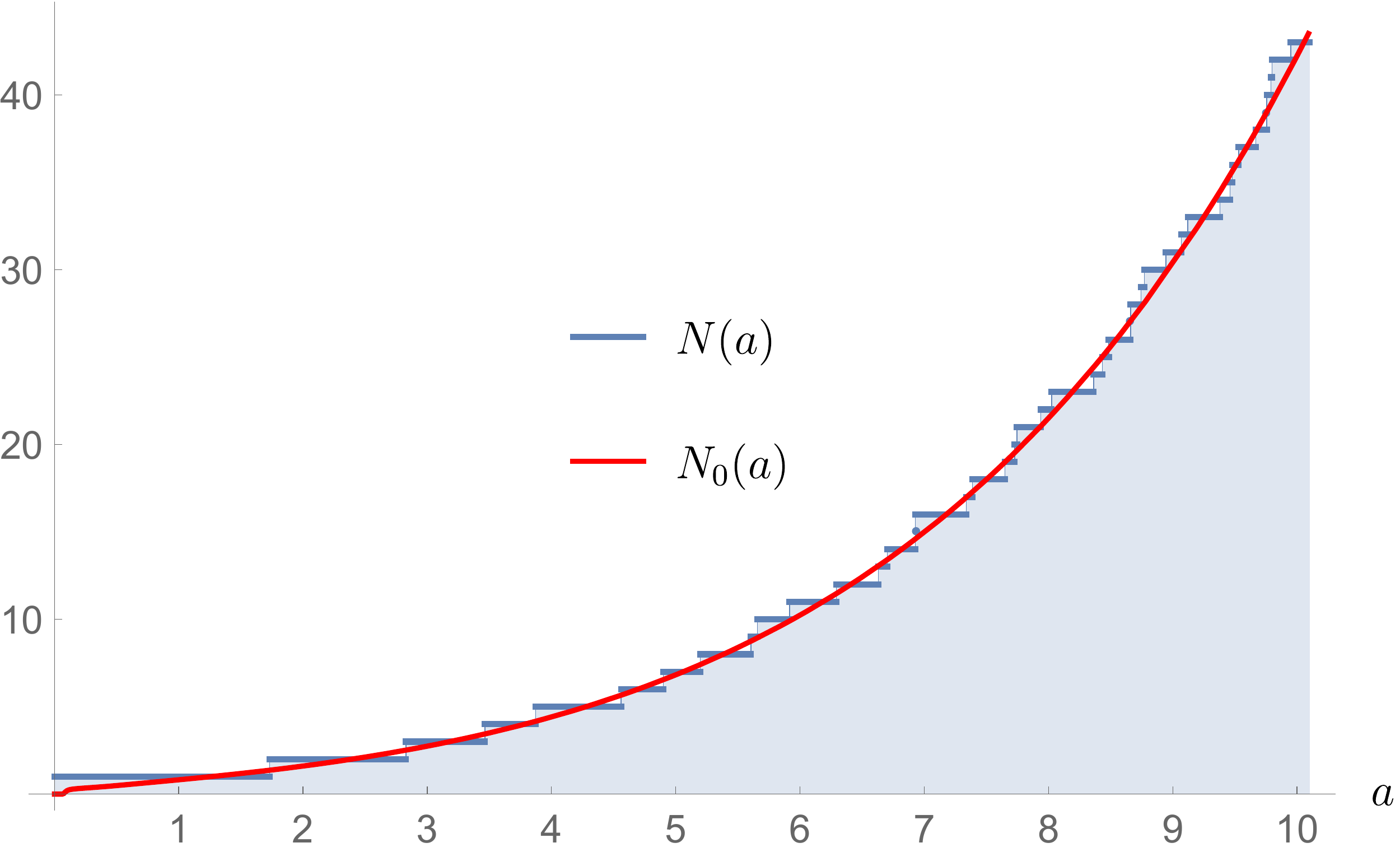}
	\caption{Comparison of the exact values of $N(a)$ and the approximation $N_0(a)$ given by \eqref{N0}. The plot obtained with the estimate \eqref{N1} is essentially identical.}\label{fig2}
\end{figure}

Let us see now how \eqref{N0} approximates the lowest part of the spectrum. To this end it suffices to plot $N(a)$ and $N_0(a)$ for small values of $a$. The result can be seen in figure \ref{fig2}. The approximations obtained above are, obviously, very good. This is in marked contrast with the one provided by \eqref{asimpt1} which, as expected, does not work very well for small values of the area.

Extending the plot for a larger range of areas will wash out the details of the staircase function in such a way that the two curves would be essentially indistinguishable. However, it is interesting to thoroughly check the quality of our approximations for large areas. In order to find out how well \eqref{N0} and \eqref{N1}, or \eqref{asimpt1} approximate $N(a)$ it is necessary to obtain exact values for $N(a)$. As $N(a)$ grows very fast this can only be done, with modest computing means, for values of $a$ of about several tens. In table \ref{Table} we give a sample of these results and compare them with the different approximations obtained in the paper.

\begin{center}
\begin{table}\centering
\begin{tabular}{ c c c c c c c c c c c }
    \hline
    & $a$  & \quad & $N_{\mathrm{as}}(a)$ & \quad & $N_0(a)$ & \quad & $N_1(a)$ & \quad & $N(a)$  & \\ \hline\hline
    & 10 & \quad & 68 	 	   & \quad & 42 	  & \quad & 42 		 & \quad & 43 	     \\ \hline
    & 20 & \quad & 979 		   & \quad & 678 	  & \quad & 679 	 & \quad & 682       \\ \hline
    & 30 & \quad & 8599 	   & \quad & 6282     & \quad & 6284 	 & \quad & 6282      \\ \hline
    & 40 & \quad & 56682 	   & \quad & 42809    & \quad & 42817 	 & \quad & 42751     \\ \hline
    & 50 & \quad & 307719 	   & \quad & 237955   & \quad & 237990   & \quad & 237788    \\ \hline
    & 60 & \quad & 1448161 	   & \quad & 1140094  & \quad & 1140222  & \quad & 1139167   \\ \hline
    & 70 & \quad & 6099037 	   & \quad & 4870521  & \quad & 4870954  & \quad & 4867770   \\ \hline
    \hline
\end{tabular}
\caption{In this table we compare exact values of $N(a)$ with the different approximations discussed in the text. We have rounded the values of all the approximations to the closest integer.  As we can see both $N_0(a)$ and $N_1(a)$ are quite accurate.}\label{Table}
\end{table}
\end{center}

\begin{figure}
  \centering
  \includegraphics[width=\textwidth]{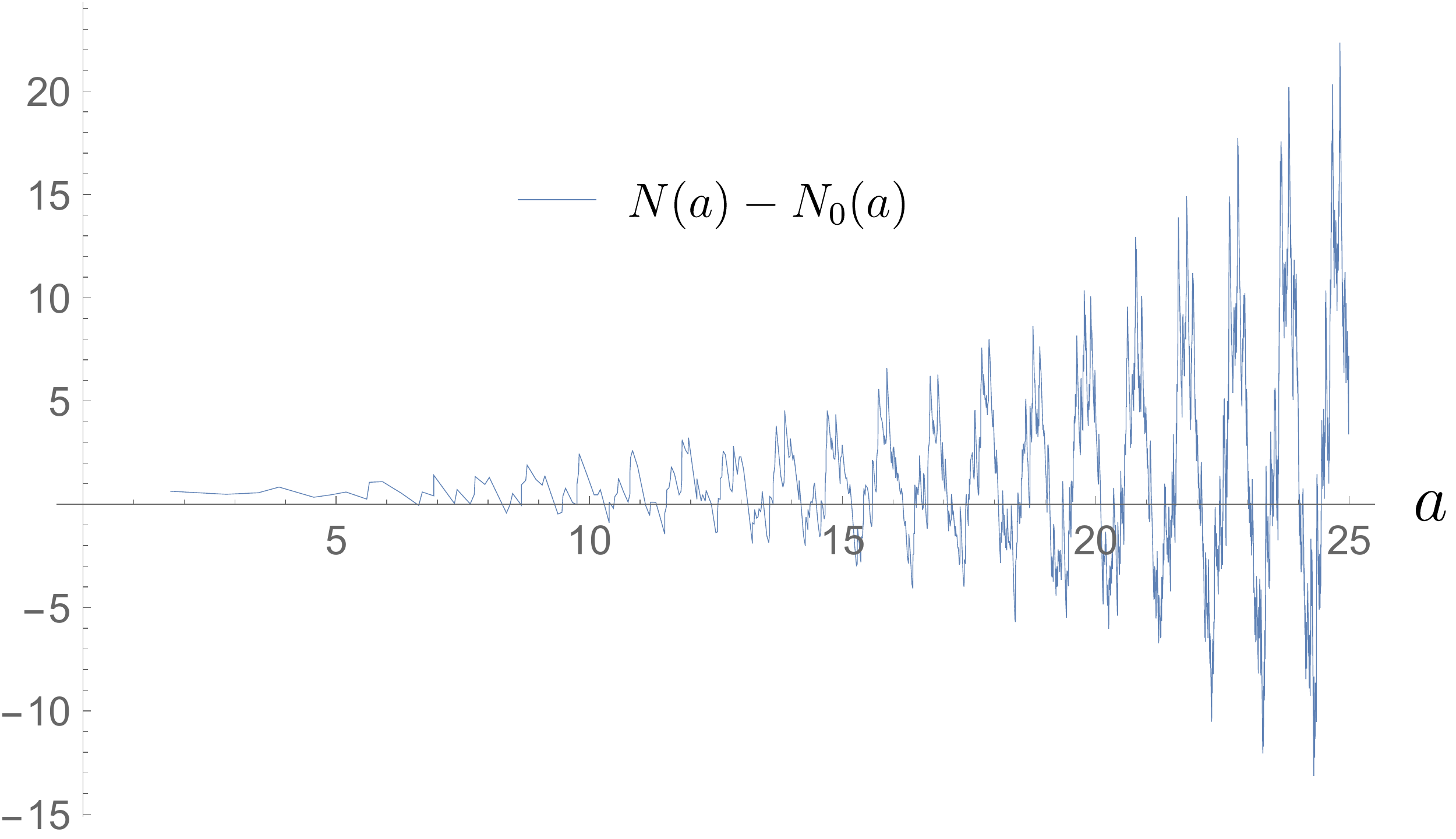}
  \caption{Plot of the difference $N(a)-N_0(a)$ evaluated only at the points of the area spectrum (we have joined the points in the plot to make it easier to understand). The exact value of $N(a)$ oscillates about the one given by $N_0(a)$. As we can see the envelope of the curve slowly increases with the area $a$. The plot with $N_1(a)$ looks essentially the same.}\label{fig3}
\end{figure}

\begin{figure}
  \centering
  \includegraphics[width=\textwidth]{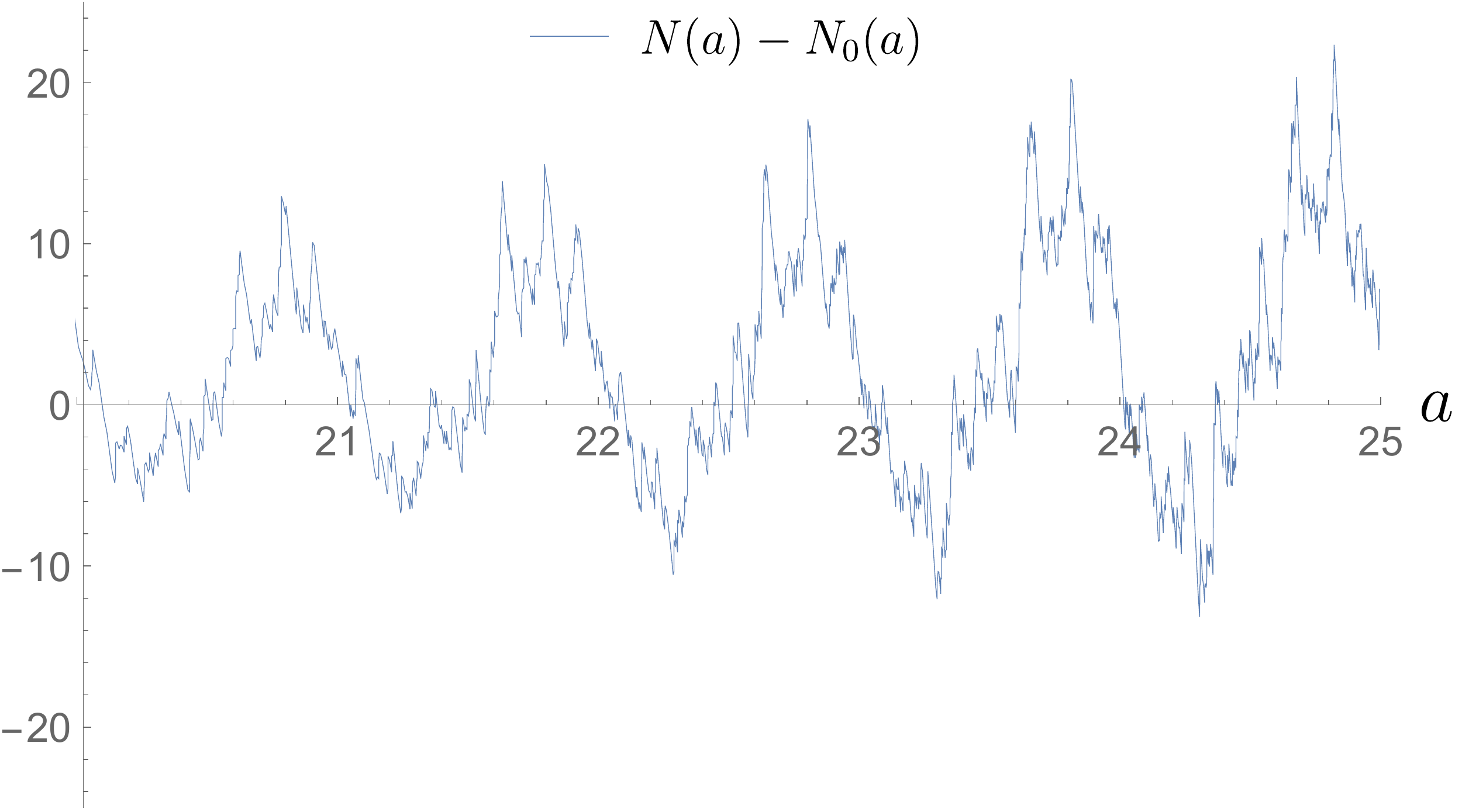}
  \caption{Zoom of the previous plot where more detail can be seen. It is interesting to notice the almost periodicity of $N(a)-N_0(a)$ once the increasing amplitude is divided out.}\label{fig4}
\end{figure}

As it can be seen, the approximations provided by $N_0(a)$ and $N_1(a)$ are increasingly good and the relative errors go to zero as the area grows (as expected). This is also true for $N_{\mathrm{as}}(a)$ although this can only be seen for very large areas.\footnote{A simple check of this fact consists in comparing the ratios of $N_0(a)$ or $N_1(a)$ with $N_{\mathrm{as}}(a)$ in the limit $a\rightarrow\infty$.} This notwithstanding, it is important to point out that the difference between $N(a)$ and $N_0(a)$ (or $N_1(a)$) does not appear to go to zero as $a\rightarrow\infty$. A good idea of what is going on can be gleaned by plotting the difference of $N(a)$ and $N_0(a)$. Two such plots are shown in figures \ref{fig3} and \ref{fig4}. As we can see $N_0(a)$ oscillates around the exact value $N(a)$ in an almost periodic fashion (the same is true for $N_1(a)$). This behaviour can be explained by the analytic structure of $\hat{f}(s)$ defined in \eqref{hatf}, in particular, the periodic concentration of singularities around the values $s=2k\pi i$ with $k\in\mathbb{Z}$. Although one can, in principle, use the methods that we have explained in the paper to explore these further corrections to $N_0(a)$ and $N_1(a)$, we think that the information that we have obtained gives a satisfactory answer to the questions that we posed at the beginning so we will stop our analysis here.

Finally, we want to point out that the quality of our approximations hinges on the fact that the lack of periodicity of $\phi(s)$ in the imaginary direction effectively makes the contributions of other saddle points negligible with respect to the one that we have used in the paper. As we will mention at the end of the appendix, this behaviour differs from the one that the analogous of $\phi(s)$ plays in the analysis of partitions.

%%%%%%%%%%%%%%%%%%%%%%%%%%%%%%%%%%%%%%%%%%%%%%%%%%%%%%%%%%%%%%%%%%%%%%%
%
% ACKNOWLEDGMENTS
%
\section*{Acknowledgments}

This work has been supported by the Spanish MINECO research grant FIS2014-57387-C3-3-P. Juan Margalef-Bentabol is supported by ``la Caixa'' and ``Residencia de Estudiantes'' fellowships. One of the authors (JFBG) wants to thank Kirill Krasnov for a discussion that prompted the analysis of the different approximations based on the partition problem discussed in the paper. We also want to thank K. Giesel, H. Sahlmann, T. Thiemann and the Quantum Gravity Group at the FAU University in Erlangen for interesting discussions and comments. Some of the computations and the plots have been done with the help of Mathematica\texttrademark.

%%%%%%%%%%%%%%%%%%%%%%%%%%%%%%%%%%%%%%%%%%%%%%%%%%%%%%%%%%%%%%%%%%%%%%%
%
% APPENDIX
%
\begin{appendices}

\section{Asymptotics of partitions via Laplace transforms}\label{appendix1}

In this appendix we will obtain the asymptotics of $p(n)$ leading to \eqref{asymptpartitions} and \eqref{asymptpartitionsmas}. Let us consider

$$N_p(n):=\sum_{k=0}^n p(k)\,,$$
and its Laplace transform representation $(n>0)$
\begin{equation}\label{Laplacepartitions}
N_p(n)=\frac{1}{2\pi i}\int_{c-i\infty}^{c+i\infty}\frac{e^{ns}}{s}\prod_{k=1}^\infty\frac{1}{1-e^{-ks}}\mathrm{d}s\,.
\end{equation}
Defining now
\begin{equation}\label{fitilde}
\phi_p(s):=-\sum_{n=1}^\infty\log(1-e^{-n s})
\end{equation}
and computing its Mellin transform
\begin{equation}\label{Mellinfitilde}
M[\phi_p;t]=\zeta(t+1)\Gamma(t)\zeta(t)\,,
\end{equation}
we get the following integral representation for $\phi_p(s)$
\begin{equation}\label{inverseMellinfitilde}
\phi_p(s)=\frac{1}{2\pi i}\int_{\hat{c}-i\infty}^{\hat{c}+i\infty}s^{-t}\zeta(t+1)\Gamma(t)\zeta(t)\mathrm{d}t\,.
\end{equation}
The integrand has simple poles at $t=+1$ and $t=-1$ and a double pole at $t=0$. By computing residues at them we get
\begin{equation}\label{expfitilde}
\exp\phi_p(s)=\sqrt{\frac{s}{2\pi}}\exp\left(\frac{\pi^2}{6s}-\frac{s}{24}\right)\cdot \exp(O(s^N))\,,\quad\forall N\in\mathbb{N}\,,s\rightarrow+\infty\,,
\end{equation}
hence
\begin{equation}\label{tildeN1}
N_p(n)=\frac{1}{(2\pi)^{3/2}i}\int_{\hat{c}-i\infty}^{\hat{c}+i\infty}
s^{-1/2}\exp{\left(\Big(n-\frac{1}{24}\Big)s+\frac{\pi^2}{6s}\right)}G(s)\mathrm{d}s\,,
\end{equation}
where $G(s)=\exp(O(s^N))$ for all $N\in\mathbb{N}$. By performing now the change of variable $s=\beta_n t$ in such a way that the real stationary point of the exponential is at $t=1$ we get
\begin{equation}\label{tildeN11}
N_p(n)=\frac{\beta_n^{1/2}}{(2\pi)^{3/2}i}\int_{\hat{c}-i\infty}^{\hat{c}+i\infty}
t^{-1/2}\exp{\left(\frac{\pi^2}{6\beta_n}\Big(t+\frac{1}{t}\Big)\right)}G(\beta_n t)\mathrm{d}t\,,
\end{equation}
with
$$
\beta_n=\frac{\pi}{\sqrt{6\Big(n-\frac{1}{24}\Big)}}\,.
$$
By proceeding as in subsection \ref{subsec_approx} we get the approximation
\begin{equation}\label{Besselpartitions}
N_p(n)\sim\frac{1}{\sqrt[4]{24n-1}}I_{-1/2}\left(\frac{\pi}{6}\sqrt{24n-1}\right)\,,
\end{equation}
leading to the asymptotic behaviour \eqref{asymptpartitionsmas}. The asymptotics of $p(n)$ can be immediately obtained by differentiating the preceding expression (or \eqref{asymptpartitionsmas}) with respect to $n$. Equation \eqref{Besselpartitions} is essentially equivalent to the first term in the convergent Rademacher expansion. It is interesting to point out that \eqref{fitilde} is periodic in the imaginary direction. It is possible then to compute the contribution of the saddle points obtained by translating the one considered here in steps of $2k\pi i$. Although these are suppressed with respect to this one owing to the $1/s$ factor in \eqref{Laplacepartitions} they will give an oscillating correction contributing to the staircase shape of $N_p(n)$.

\end{appendices}

\end{document}